\documentclass[manuscript,screen,nonacm]{acmart}
\AtBeginDocument{%
  }

\setcopyright{acmlicensed}
\copyrightyear{2025}
\acmYear{2025}
\acmDOI{XXXXXXX.XXXXXXX}
\acmConference[SCA/HPCAsia '26]{SCA-HPCAsia}{January 26--29, 2026}{Osaka, Japan}
\acmISBN{978-1-4503-XXXX-X/2018/06}

\usepackage{booktabs,tabularx}
\usepackage{quantikz}
\usepackage{braket} 
\usepackage{soul}

\usepackage{siunitx}
\usepackage[most]{tcolorbox}
\tcbset{
  colback=gray!3, 
  colframe=black,
  boxrule=0.4pt,
  sharp corners,
  left=1mm,right=1mm,top=1mm,bottom=1mm
}
\newtcolorbox{defbox}[1]{title={#1}, breakable, fonttitle=\bfseries}

\begin{document}
 
\title{QPU Micro-Kernels for Stencil Computation}

\author{Stefano Markidis}
\affiliation{%
  \institution{KTH Royal Institute of Technology}
  \city{Stockholm}
  \country{Sweden}}

  \author{Luca Pennati}
\affiliation{%
  \institution{KTH Royal Institute of Technology}
  \city{Stockholm}
  \country{Sweden}}

  \author{Marco Pasquale}
\affiliation{%
  \institution{KTH Royal Institute of Technology}
  \city{Stockholm}
  \country{Sweden}}

\author{Gilbert Netzer}
\affiliation{%
  \institution{KTH Royal Institute of Technology}
  \city{Stockholm}
  \country{Sweden}}

\author{Ivy Peng}
\affiliation{%
  \institution{KTH Royal Institute of Technology}
  \city{Stockholm}
  \country{Sweden}}

\renewcommand{\shortauthors}{Markidis et al.}

\begin{abstract}
We introduce QPU micro-kernels: shallow quantum circuits that perform a stencil node update and return a Monte Carlo estimate from repeated measurements. We show how to use them to solve Partial Differential Equations (PDEs) explicitly discretized on a computational stencil. From this point of view, the QPU serves as a sampling accelerator. Each micro-kernel consumes only stencil inputs (neighbor values and coefficients), runs a shallow parameterized circuit, and reports the sampled mean of a readout rule. The resource footprint in qubits and depth is fixed and independent of the global grid. This makes micro-kernels easy to orchestrate from a classical host and to parallelize across grid points. We present two realizations. The Bernoulli micro-kernel targets convex-sum stencils by encoding values as single-qubit probabilities with shot allocation proportional to stencil weights. The branching micro-kernel prepares a selector over stencil branches and applies addressed rotations to a single readout qubit. In contrast to monolithic quantum PDE solvers that encode the full space-time problem in one deep circuit, our approach keeps the classical time loop and offloads only local updates. Batching and in-circuit fusion amortize submission and readout overheads. We test and validate the QPU micro-kernel method on two PDEs commonly arising in scientific computing: the Heat and viscous Burgers’ equations. On noiseless quantum circuit simulators, accuracy improves as the number of samples increases. On the IBM \texttt{Brisbane} quantum computer, single-step diffusion tests show lower errors for the Bernoulli realization than for branching at equal shot budgets, with QPU micro-kernel execution dominating the wall time.
\end{abstract}



\maketitle

\section{Introduction}
Stencil computations are a cornerstone of scientific computing. They arise from the discretization of Partial Differential Equations~(PDEs) on numerical grids~\cite{farlow1993partial,morton2005numerical,datta2008stencil}. At each solver iteration, the value at a grid point is updated from a fixed pattern of neighboring points, known as the \emph{stencil}. In explicit discretizations, the update at time level $n+1$ depends only on values at level $n$, so each point can be updated independently and in an embarrassingly parallel fashion. This property maps well to GPUs, where a host launches a device kernel that performs the per point update concurrently across threads~\cite{kirk2016programming,holewinski2012high}. Motivated by this model, we introduce QPU kernels that implement the same per-point stencil update on a quantum processor and can be composed and scheduled within heterogeneous classical-quantum solvers.

Quantum computing is an emerging computational model that exploits superposition, interference, and entanglement to process information~\cite{nielsen2010quantum,feynman2018feynman,markidis2024quantum}. Quantum computing systems have advanced to devices with hundreds to thousands of physical qubits. From an HPC perspective, QPUs can be abstracted as specialized accelerators that execute parameterized quantum circuits launched from a classical host~\cite{bertels2020quantum}. In most workflows, these purely quantum kernels, such as a Harrow-Hassidim-Lloyd~(HHL) linear solver~\cite{harrow2009quantum}, are \emph{monolithic} circuits tailored to a complete problem instance, which increases circuit depth and resource usage~\cite{hegde2024beyond}. In contrast, in this work we adopt a kernel-style formulation analogous to GPUs and use the \ul{QPU as a sampling accelerator}. We cast the per-iteration update at a single grid point as a small quantum subroutine. This leads to circuits with a minimal qubit footprint and shallow depth that are compatible with current Noisy Intermediate-Scale Quantum~(NISQ)~\cite{preskill2017quantum} processors. We refer to these circuits as \emph{micro-kernels}. The construction follows a Monte Carlo approach~\cite{caflisch1998monte,robert1999monte}. PDE coefficients and previous-time-level values are encoded as probabilities via a Bernoulli encoder, and a composition of single-qubit rotation and controlled-rotation gates produces a readout qubit measurement probability equal to the stencil update. 

\begin{figure}[t]
\centering
    \hspace{0mm}
    \begin{minipage}{0.225\linewidth}
        \centering
        \caption*{\centering (a) Monolithic Quantum \\ PDE Solver}

        \vspace{0.6mm}
        \includegraphics[width=\linewidth, clip = true, trim = 0 0 12cm 0]{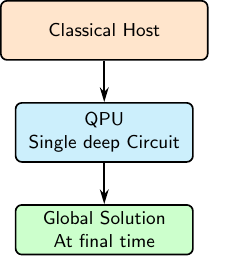}
    \end{minipage}
    \begin{minipage}{0.72\linewidth}
        \centering
        \caption*{\hspace{-1cm}(b) PDE Solver with QPU Micro-Kernels}

        \vspace{3.4mm}
        \includegraphics[width=\linewidth, clip = true, trim = 3.6cm 0 0 0]{QPU_tickz_graph.pdf}
    \end{minipage}    
    \caption{Comparison of traditional quantum PDE solvers and the proposed QPU micro-kernel method.}
    \label{fig:diagram}
\end{figure}

The monolithic quantum solver and the QPU micro-kernel paradigms have different purposes and target different platforms. Fig.~\ref{fig:diagram} compares them. In panel~(a), a monolithic quantum PDE solver encodes the entire problem in a single deep circuit and aims to produce the global field at the final time in one shot. Circuit width, depth, and compilation complexity grow with grid size and time span, stressing current NISQ quantum hardware. In panel~(b), our approach, a micro-kernel-based solver, keeps the classical time loop and offloads only the node-local stencil update to shallow circuits. The QPU micro-kernel method scales independently of the global grid and can be run in parallel across nodes. The micro-kernel model is a good fit for explicit schemes on current NISQ hardware. By contrast, monolithic quantum circuits target global formulations in a fully fault-tolerant regime~\cite{nielsen2010quantum}.

The goal of this paper is to investigate QPUs as sampling accelerators for explicit stencil-based PDE solvers. We design QPU micro-kernel circuit realizations, analyze their accuracy and performance, and demonstrate their use on representative 1D Heat and viscous Burgers’ equations. Our contributions are the following:
\begin{itemize}
  \item We introduce \emph{QPU micro-kernels}. These are shallow circuits that perform a single explicit stencil update by encoding neighbor values and weights with $R_y$ rotations so that a readout qubit’s success probability equals the finite-difference update.
  \item We develop and explain in detail two QPU micro-kernel implementations: \emph{(i)} a quantum Bernoulli micro-kernel that mixes one-qubit encoders with shot allocation proportional to stencil weights, and \emph{(ii)} a quantum branching micro-kernel that prepares a selector register for categorical weights and applies leaf-addressed rotations to a single readout qubit. We also outline additional variants, including \emph{signed-mixture} micro-kernels for updates with signed coefficients and \emph{noise} micro-kernels for stochastic problems. We propose and analyze \emph{in-circuit micro-kernel fusion} and \emph{batched submission} to amortize host-device launch and readout overheads.
  \item We implement the approach for two PDEs: the 1D Heat equation and the 1D viscous Burgers’ equation. We provide results using a noiseless quantum circuit simulator and on real quantum hardware. We present a runtime breakdown showing that device execution dominates per-job wall time, and we quantify transpilation, queue, and execution costs for one solver iteration.
\end{itemize}

\section{Monte Carlo Realization of Stencil Updates}
We focus on two canonical PDEs used for modeling transport and diffusion phenomena in science and engineering: the 1D Heat and viscous Burgers' equations.

\subsection{One-dimensional Heat Equation}
As first PDE, we consider a representative parabolic PDE for how a quantity (temperature, concentration, etc.) spreads in space. We use $u(x,t)$ to denote that quantity at position $x$ and time $t$. On the unit interval, with constant diffusivity $\nu>0$, the one-dimensional Heat (also called \emph{diffusion}) equation with a source $s(x,t)$ is
\begin{equation}
\label{eq:heat}
\partial_t u(x,t) \;=\; \nu\,\partial_{xx} u(x,t) \;+\; s(x,t),
\qquad x\in(0,1),\ t>0,
\end{equation}
together with homogeneous Dirichlet boundary conditions $u(0,t)=0, u(1,t)=0$, and an initial condition $u(x,0)=u_0(x)$. For validation, we often choose $u_0(x)=\sin(\pi x)$ and $s\equiv 0$, which admits the closed-form solution as $u(x,t)=e^{-\nu\pi^2 t}\sin(\pi x)$.

To evolve this dynamics forward in time, we place $N$ interior points on a uniform grid
\begin{equation}
x_i=i\,\Delta x,\qquad \Delta x=\frac{1}{N+1},\qquad i=0,1,\dots,N{+}1,
\end{equation}
so that $x_0=0$ and $x_{N+1}=1$ are the boundaries. We advance at discrete times $t^n=n\,\Delta t$. We write $u_i^n\approx u(x_i,t^n)$ and enforce the boundaries directly by fixing $u_0^n=u_{N+1}^n=0$ for all $n$. Using a centered second order difference for $\partial_{xx}$ and a forward Euler step for $\partial_t$ gives the familiar explicit Forward Time-Centered Space (FTCS) update, valid for interior indices $i=1,\dots,N$:
\begin{equation}
\label{eq:ftcs}
u_i^{\,n+1}
\;=\;
u_i^{\,n}
\;+\;
\lambda\big(u_{i-1}^{\,n}-2u_i^{\,n}+u_{i+1}^{\,n}\big)
\;+\;
\Delta t\,s(x_i,t^n),
\qquad
\lambda=\frac{\nu\,\Delta t}{\Delta x^2}.
\end{equation}
This formula can be easily understood. Diffusion leads $u_i$ toward the average of its neighbors. Rearranging the diffusion term makes this explicit,
\begin{equation}
\label{up_dif}
u_i^{\,n+1}
\;=\;
\lambda\,u_{i-1}^{\,n} \;+\; (1-2\lambda)\,u_i^{\,n} \;+\; \lambda\,u_{i+1}^{\,n}
\;+\; \Delta t\,s(x_i,t^n).
\end{equation}
The update corresponds to a convex combination (a weighted average with nonnegative coefficients summing to 1), so the result lies between the smallest and largest of $u_{i-1},u_i,u_{i+1}$ with weights $(\lambda,\,1-2\lambda,\,\lambda)$, respectively. Intuitively, each node update reduces to computing that weighted average and, if present, adding the explicit source. Finally, the explicit scheme is stable, provided the time step is small enough relative to the mesh spacing and diffusivity. On a uniform grid with constant $\nu$ the standard condition is $0 \;\le\; \lambda \;\le\; \tfrac12.$

\ul{Eq.}~\ref{up_dif} \ul{can be viewed as a one-step \emph{random walk}}. From node $i$ we move to the left ($i{-}1$) with probability $\lambda$, stay at $i$ with
probability $1-2\lambda$, and move to the right ($i{+}1$) with probability $\lambda$. We denote this three-point choice by a categorical random variable
$X$ that takes values in $\{i{-}1,i,i{+}1\}$ with probabilities $(\lambda,1-2\lambda,\lambda)$. We write this compactly as
$$
X \sim \mathrm{Cat}(i{-}1{:}\lambda,\; i{:}1-2\lambda,\; i{+}1{:}\lambda),
$$
meaning that $X$ returns $i{-}1$ with probability $\lambda$, $i$ with probability $1-2\lambda$, and $i{+}1$ with probability $\lambda$. With this shorthand, the FTCS update is simply the expected value at the chosen location plus the explicit source,
$$
u_i^{\,n+1}
\;=\;
\mathbb{E}\!\big[\,u_{X}^{\,n}\,\big] \;+\; \Delta t\,s(x_i,t^n).
$$

In practice, sampling this three-point distribution is straightforward and can be implemented with a single uniform random number and two thresholds. If we
repeat the random choice $M$ times and average the sampled values, we obtain the Monte Carlo estimator~\cite{caflisch1998monte,robert1999monte}:
\begin{equation}
\label{eq:mc-estimator-simple}
\widehat{u}_i^{\,n+1}
\;=\;
\frac{1}{M}\sum_{m=1}^M u_{X^{(m)}}^{\,n}
\;+\; \Delta t\, s(x_i,t^n),
\qquad
X^{(m)} \sim \mathrm{Cat}(i{-}1{:}\lambda,\; i{:}1-2\lambda,\; i{+}1{:}\lambda)\ \text{i.i.d.}
\end{equation}
which approximates the deterministic update as $M$ grows.

\subsection{One-Dimensional Viscous Burgers’ Equation}
As a second PDE, we consider the one-dimensional viscous Burgers’ equation in conservative form:
\begin{equation}
\partial_t u(x,t)\;+\;\partial_x f\!\big(u(x,t)\big)\;=\;\nu\,\partial_{xx}u(x,t),
\qquad f(u)=\tfrac12 u^2,\qquad x\in[-1,1],\ t\in[0,1],
\label{eq:burgers-pde}
\end{equation}
equivalently $\partial_t u+u\,\partial_x u=\nu\,\partial_{xx}u$. We impose homogeneous Dirichlet boundaries $u(t,\pm1)=0$, the initial condition $u(0,x)=-\sin(\pi x)$, and viscosity $\nu=0.01/\pi$. On a uniform grid $x_i=-1+i\,\Delta x$ with $\Delta x=2/(N{+}1)$ and time levels $t^n=n\,\Delta t$, we write $u_i^n\approx u(x_i,t^n)$ for $i=1,\dots,N$ and enforce boundaries by ghost cells $u_0^n=u_{N+1}^n=0$. We advance $u_i$ by a convex average of neighbors with node-local weights that depend on $u_i^n$ and $\nu$:
\begin{equation}
u_{i}^{\,n+1} \;=\; w_L\,u_{i-1}^{\,n} \;+\; w_C\,u_{i}^{\,n} \;+\; w_R\,u_{i+1}^{\,n},
\qquad
w_L,w_C,w_R\ge 0,\quad w_L{+}w_C{+}w_R=1,
\label{eq:burgers-convex}
\end{equation}
with
\begin{equation}
c=\frac{u_i^{\,n}\,\Delta t}{\Delta x},\qquad
c^{+}=\max(c,0),\ \ c^{-}=\max(-c,0),\qquad
w_L=\lambda +c^{+},\ \ w_R=\lambda +c^{-},\ \ w_C=1-(w_L{+}w_R).
\label{eq:burgers-weights}
\end{equation}
The split $c^{\pm}=\max(\pm\,u_i^n\,\Delta t/\Delta x,0)$ is the standard \textit{Courant splitting} for a first-order \textit{upwind} discretization of the nonlinear advection $u\,u_x$. Adding the centered Laplacian term $\lambda(u_{i-1}^n-2u_i^n+u_{i+1}^n)$ (with the same diffusive CFL parameter $\lambda=\nu\Delta t/\Delta x^2$ as in Eq.~\ref{eq:ftcs}) leads to the explicit \emph{upwind\,+\,FTCS} step in Eqs.~\ref{eq:burgers-convex}-\ref{eq:burgers-weights}. In conservative flux form this is equivalent, for scalar laws, to a \emph{Godunov} update with interface upwinding (or, equivalently, a local Lax-Friedrichs/Rusanov flux for the advective part); see, for instance, Ref.~\cite{leveque2002finite}. It is important to note that this form is monotone under the combined advective-diffusive Courant-Friedrichs-Lewy (CFL) 
\begin{equation}
\frac{|u_i^{\,n}|\,\Delta t}{\Delta x} \;+\; 2\,\frac{\nu\,\Delta t}{\Delta x^2} \;\le\; 1
\quad\text{for all } i,
\label{eq:burgers-cfl}
\end{equation}
which guarantees $w_L,w_C,w_R\ge 0$ and $w_L{+}w_C{+}w_R=1$. Eq.~\ref{eq:burgers-convex} is an expectation over a three-point categorical draw. If
$
X_i^n\ \sim\ \mathrm{Cat}\!\big(i{-}1{:}w_L,\ i{:}w_C,\ i{+}1{:}w_R\big),
$
then
\begin{equation}
u_i^{\,n+1} \;=\; \mathbb{E}\!\big[\,u_{X_i^n}^{\,n}\,\big],
\qquad
\widehat{u}_i^{\,n+1} \;=\; \frac{1}{M}\sum_{m=1}^{M} u_{X_i^{n,(m)}}^{\,n},
\quad X_i^{n,(m)}\ \sim\ \mathrm{Cat}\!\big(i{-}1{:}w_L,\ i{:}w_C,\ i{+}1{:}w_R\big) \ \text{with independent draws}.
\label{eq:burgers-mc-estimator}
\end{equation}

\section{Quantum Micro-Kernel Method}
We reformulate each explicit stencil update as the expectation of a local random variable. We evaluate it with quantum subroutines that we call \emph{QPU micro-kernels}. Each kernel encodes stencil neighbor values and coefficients, executes a shallow circuit, and returns a Monte Carlo estimate of the desired scalar update (the unknown value at the node position). This section defines the abstraction and then it presents two main circuit realizations, followed by the description of other micro-kernels, extension to higher dimensions, batching and fusion techniques, and a performance model.
\begin{defbox}{}
A QPU micro-kernel is a shallow parameterized circuit $U(x;\theta)$ on $q$ qubits that, for classical inputs $x\in\mathbb{R}^k$ and parameters $\theta$, prepares
$$
\ket{\psi(x;\theta)} \;=\; U(x;\theta)\ket{\mathbf{0}},
$$
where $\ket{\mathbf{0}}$ is the all-zero computational basis state. A single measurement of $\ket{\psi(x;\theta)}$ produces a bitstring $B\in\{0,1\}^q$ with distribution $P_{x,\theta}$.
We associate a scalar to that outcome via a readout rule $r:\{0,1\}^q\to\mathbb{R}$.
The quantity of interest is the mean of this scalar under the measurement distribution. Given $M$ measurements,
$$
z(x;\theta) \;=\; \mathbb{E}_{B\sim P_{x,\theta}}\!\big[\,r(B)\,\big],
\qquad
\widehat{z}(x;\theta) \;=\; \frac{1}{M}\sum_{s=1}^{M} r\big(B^{(s)}\big).
$$
Each invocation is independent and has a fixed, small resource footprint (number of qubits $q$ and circuit depth $d$) suitable for batching.
\end{defbox}

We note that \ul{QPU micro-kernels without the readout can be viewed as feature maps} used in Quantum Machine Learning (QML)~\cite{schuld2021supervised,markidis2023programming}. If we fix a small set of observables on the circuit, then each call to the micro-kernel maps the stencil input $x$ to a short vector of expectation values (one entry per observable). A per-node update can be seen as a linear combination of the entries of this vector. This point of view defines a kernel function over stencil inputs. The similarity between two inputs $x$ and $x'$ is given by the inner product of their expectation vectors. Collecting these similarities into a matrix (the associated Gram or kernel matrix) allows us to apply standard spectral tools~\cite{scholkopf2002learning,andrew2000introduction}. For example, we can inspect eigenvalue decay or effective rank to study the numerical properties of the QPU micro-kernel method and to evaluate more advanced variants such as higher-order stencils or adaptive sampling.

\subsection{QPU Micro-Kernel Realization}
We now describe the circuit design of QPU micro-kernels. We first present two realizations for explicit three-point stencils, the Bernoulli and branching micro-kernels, and then outline additional variants (signed-mixture, row, and noise micro-kernels) that cover common stencil and linear-algebra operations.

\subsubsection{Bernoulli Micro-Kernel}
The first quantum micro-kernel implements a circuit to encode a scalar value $u\in[0,1]$ into a qubit amplitude. We implement a unitary map from $u$ to a single-qubit state whose computational-basis measurement produces $1$ with probability $u$ (Fig.~\ref{fig:bernoulli-encoder}). The qubit is initialized in $\ket{0}$. For any rotation angle $\theta\in\mathbb{R}$,
\begin{equation}
R_y(\theta)\ket{0}
= \cos\!\Big(\frac{\theta}{2}\Big)\ket{0}
+ \sin\!\Big(\frac{\theta}{2}\Big)\ket{1}.
\label{eq:ry-on-zero}
\end{equation}
The probability of outcome $1$ is $\Pr(1)=\sin^2(\theta/2)$. Choosing
\begin{equation}
\theta(u) \;=\; 2\,\arcsin\sqrt{u}
\qquad\text{(equivalently, }2\,\arccos\sqrt{1-u}\text{)}
\label{eq:bernoulli-angle}
\end{equation}
prepares $\ket{\psi_u}=R_y(\theta(u))\ket{0}$ with $\Pr(1)=u$, i.e., a Bernoulli$(u)$ encoder.

\begin{figure}[h!]
  \centering
  \begin{quantikz}
    \lstick{$\text{ro}$: $|0\rangle$} & \gate{R_y\!\big(2\arcsin\sqrt{u}\big)} & \qw
  \end{quantikz}
  \caption{Bernoulli encoder: a single $R_y$ rotation maps $u\in[0,1]$ to a qubit measured $1$ with probability $u$. After the gate the state is
  $\cos(\tfrac{\theta}{2})|0\rangle+\sin(\tfrac{\theta}{2})|1\rangle$ with
  $\theta=2\arcsin\sqrt{u}$.}
  \label{fig:bernoulli-encoder}
\end{figure}
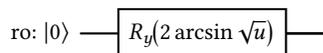
\begin{defbox}{}
The constructions above assume a normalized scalar input $u\in[0,1]$, which is interpreted as the success probability of a Bernoulli encoder. When the
physical field takes values in a different range $[u_{\min},u_{\max}]$, we apply an affine normalization $v=(u-u_{\min})/(u_{\max}-u_{\min})\in[0,1]$ before preparing the quantum state, and map the estimator back via $u = u_{\min} + (u_{\max}-u_{\min})\,v$. 
\end{defbox}
To solve Eq.~\ref{up_dif} with a \ul{quantum Bernoulli micro-kernel}, we build $z_i$ as a shot-weighted average of three independent one-qubit encoders (branches $b\in\{\mathrm{L},\mathrm{C},\mathrm{R}\}$). Given a total of $M$ shots for node $i$, allocate
\begin{equation}
M_{\mathrm{L}}=\big\lfloor \lambda M \big\rceil,\quad
M_{\mathrm{C}}=\big\lfloor (1-2\lambda) M \big\rceil,\quad
M_{\mathrm{R}}=M-M_{\mathrm{L}}-M_{\mathrm{C}}.
\label{eq:shot-alloc}
\end{equation}
For each branch value $u_b\in[0,1]$, we prepare $R_y\!\big(2\arcsin\sqrt{u_b}\big)\ket{0}$, measure $M_b$ times to obtain the empirical fraction $\hat u_b$ of outcomes $1$, and use these branch estimates in the convex combination. The node estimator is then
\begin{equation}
\hat z_i \;=\; \sum_{b\in\{\mathrm{L},\mathrm{C},\mathrm{R}\}} \frac{M_b}{M}\,\hat u_b
\;=\; \frac{M_{\mathrm{L}}\hat u_{\mathrm{L}} + M_{\mathrm{C}}\hat u_{\mathrm{C}} + M_{\mathrm{R}}\hat u_{\mathrm{R}}}{M}.
\label{eq:cm-estimator}
\end{equation}

\begin{defbox}{}
\paragraph{Spatial sampling error.}
With the Bernoulli encoder, a local value $u\in[0,1]$ is represented as the success probability of a single-qubit measurement. After $M$ shots, the estimator $\hat u$ is the sample mean of $M$ independent Bernoulli$(u)$ outcomes. Its standard error is
$$
\mathrm{SE}(\hat u)
\;=\; \sqrt{\frac{u(1-u)}{M}}
\;\le\; \frac{1}{2\sqrt{M}}.
$$
This follows from the binomial variance of the sample mean, $\operatorname{Var}(\hat u)=u(1-u)/M$. In the Bernoulli encoder, the sampling error is largest for mid-range values $u\approx\tfrac12$ and smallest near $u\approx0$ or $u\approx1$.
\end{defbox}

\subsubsection{Branching Micro-Kernel}
The branching micro-kernel uses superposition to implement the selection among multiple branches (categories) inside a single quantum circuit. We build it
from four simple ingredients: a conditional injector, a binary selector, a multiway selector, and leaf-addressed value loading.

\paragraph{Step 1: Conditional injector.}
We first implement the law of total probability on a single readout qubit. Let $P$ be a \emph{parent} qubit and $A$ the readout. We want the marginal $\Pr(A{=}1) \;=\; \Pr(P{=}1)\,p_1 \;+\; \Pr(P{=}0)\,p_0$, where $p_1$ and $p_0$ are user-specified conditional probabilities. As in the Bernoulli micro-kernel, a rotation $R_y(\phi)$ on $\ket{0}$ encodes a Bernoulli random variable with mean $\sin^2(\phi/2)$. Therefore, we choose angles $\phi_1$ and $\phi_0$ so that $\sin^2(\phi_b/2)=p_b$ for $b\in\{0,1\}$. A controlled $R_y(\phi_1)$ on $A$ that fires when $P{=}1$ injects the $P{=}1$ branch. A second controlled $R_y(\phi_0)$, implemented with a negative control via an $X$-control-$X$ sandwich on $P$, injects the $P{=}0$ branch. Together they realize the desired marginal on $A$.
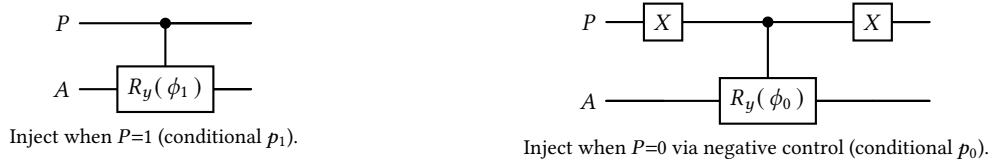
\begin{figure}[h!]
  \centering
  \begin{minipage}{0.47\linewidth}
    \centering
    \begin{quantikz}
      \lstick{$P$} & \ctrl{1} & \qw \\
      \lstick{$A$} & \gate{R_y(\,\phi_1\,)} & \qw
    \end{quantikz}

    \vspace{2pt}
    {\small Inject when $P{=}1$ (conditional $p_1$).}
  \end{minipage}
  \hfill
  \begin{minipage}{0.47\linewidth}
    \centering
    \begin{quantikz}
      \lstick{$P$} & \gate{X} & \ctrl{1} & \gate{X} & \qw \\
      \lstick{$A$} & \qw      & \gate{R_y(\,\phi_0\,)} & \qw & \qw
    \end{quantikz}

    \vspace{2pt}
    {\small Inject when $P{=}0$ via negative control (conditional $p_0$).}
  \end{minipage}
  \caption{Conditional injector: two controlled $R_y$ rotations load the conditionals $p_1$ and $p_0$ on the readout $A$.}
  \label{fig:p2-conditional-injector}
\end{figure}

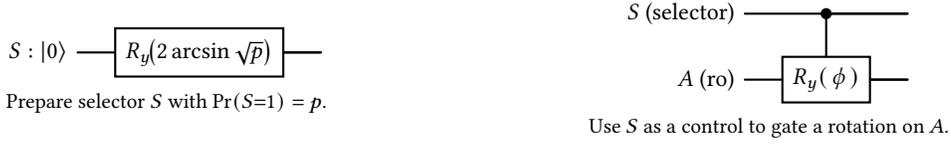
\begin{figure}[h!]
  \centering
  \begin{minipage}{0.47\linewidth}
    \centering
    \begin{quantikz}
      \lstick{$S:\ket{0}$} & \gate{R_y\!\big(2\arcsin\sqrt{p}\big)} & \qw
    \end{quantikz}

    \vspace{2pt}
    {\small Prepare selector $S$ with $\Pr(S{=}1)=p$.}
  \end{minipage}
  \hfill
  \begin{minipage}{0.47\linewidth}
    \centering
    \begin{quantikz}
      \lstick{$S$ (selector)} & \ctrl{1} & \qw \\
      \lstick{$A$ (ro)}       & \gate{R_y(\,\phi\,)} & \qw
    \end{quantikz}

    \vspace{2pt}
    {\small Use $S$ as a control to gate a rotation on $A$.}
  \end{minipage}
  \caption{Quantum binary selector: a single $R_y$ prepares $S$, which is then used as a control to weight downstream operations.}
  \label{fig:p3-selector}
\end{figure}

\paragraph{Step 2: Binary selector.} Next we prepare a \emph{selector} qubit $S$ whose outcomes realize a prescribed Bernoulli distribution $\Pr(S{=}1)=p$, as shown in Fig.~\ref{fig:p3-selector}. Using Eq.~\ref{eq:ry-on-zero}, we choose a rotation angle $\theta(p)$ so that measuring $S$ in the computational basis would produce $1$ with probability $p$ and $0$ with probability $1-p$. In the branching micro-kernel, $S$ is not measured. Instead, it acts as a control line that splits amplitude into two branches with mixing weights $p$ and $1-p$.

\begin{figure}[h!]
  \centering
  \begin{quantikz}
    \lstick{$s_0:\ket{0}$} & \gate{R_y(\theta_0)} & \gate{X} & \ctrl{1} & \gate{X} & \ctrl{1} & \qw \\
    \lstick{$s_1:\ket{0}$} & \qw                   & \qw      & \gate{R_y(\theta_L)} & \qw & \gate{R_y(\theta_R)} & \qw
  \end{quantikz}
  \caption{Four-way selector: a rotation on $s_0$ sets left/right weight, and controlled rotations on $s_1$ split each side into two leaves.}
  \label{fig:p4-rotation-tree}
\end{figure}
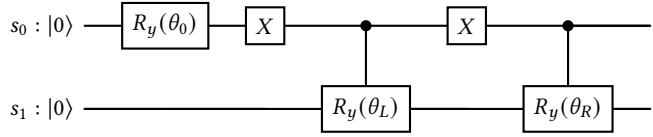

\paragraph{Step 3: Multiway selector.} To handle more than two branches, we combine several selector qubits in a small rotation tree. The idea is to factor a categorical distribution along a binary tree. Each internal node uses a single $R_y$ rotation to split its total mass into \textit{left} and \textit{right} children, and we perform recursion down the tree until each leaf corresponds to one branch. For example, a four-way selector can be built with two selector qubits
$s_0$ and $s_1$, as shown in Fig.~\ref{fig:p4-rotation-tree}. A rotation on $s_0$ sets the total weight of the left vs.\ right half of the tree, and
controlled rotations on $s_1$ further split each half into two leaves. Negative controls are again implemented with $X$-control-$X$ sandwiches. The resulting
leaf statistics leads to the desired four-way probabilities.

\begin{figure}[t]
  \centering
  \begin{minipage}{0.32\linewidth}
    \centering
    \begin{quantikz}
      \lstick{$s_0$}   & \gate{X} & \ctrl{2} & \gate{X} & \qw \\
      \lstick{$s_1$}   & \gate{X} & \ctrl{1} & \gate{X} & \qw \\
      \lstick{$\text{ro}$} & \qw & \gate{R_y(\phi_{00})} & \qw & \qw
    \end{quantikz}

    \vspace{2pt}
    {\small Leaf $00$: match $s_0{=}0$, $s_1{=}0$ and apply $R_y(\phi_{00})$ on ro.}
  \end{minipage}
  \hfill
  \begin{minipage}{0.32\linewidth}
    \centering
    \begin{quantikz}
      \lstick{$s_0$}   & \gate{X} & \ctrl{2} & \gate{X} & \qw \\
      \lstick{$s_1$}   & \qw      & \ctrl{1} & \qw      & \qw \\
      \lstick{$\text{ro}$} & \qw & \gate{R_y(\phi_{01})} & \qw & \qw
    \end{quantikz}

    \vspace{2pt}
    {\small Leaf $01$: match $s_0{=}0$, $s_1{=}1$ and apply $R_y(\phi_{01})$.}
  \end{minipage}
  \hfill
  \begin{minipage}{0.32\linewidth}
    \centering
    \begin{quantikz}
      \lstick{$s_0$}   & \qw & \ctrl{2} & \qw & \qw \\
      \lstick{$s_1$}   & \qw & \qw      & \qw & \qw \\
      \lstick{$\text{ro}$} & \qw & \gate{R_y(\phi_{10})} & \qw & \qw
    \end{quantikz}

    \vspace{2pt}
    {\small Leaf $10$: match $s_0{=}1$ (leaf $11$ unused) and apply $R_y(\phi_{10})$.}
  \end{minipage}
  \caption{Leaf loading: each panel applies a rotation $R_y(\phi_\ell)$ to the
  readout only when the selector $(s_0 s_1)$ matches leaf $\ell$.}
  \label{fig:p5-leaf-injection}
\end{figure}
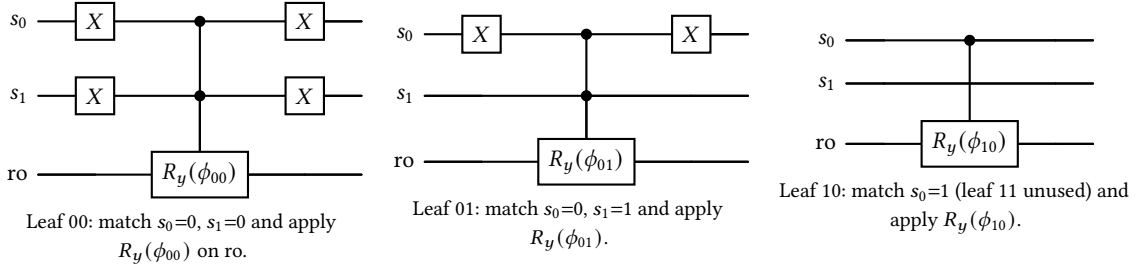

\paragraph{Step 4: Loading branch values.} The last ingredient is to load branch values onto a single readout qubit. Once the selector qubits have been prepared so that each leaf is visited with the desired probability, we associate a value $u_\ell\in[0,1]$ to each leaf $\ell$ and encode it via a rotation $R_y(\phi_\ell)$ on the readout, with $\sin^2(\phi_\ell/2)=u_\ell$. Each $R_y(\phi_\ell)$ is applied only on the subspace where the selector bits match the corresponding leaf bit pattern. After all nonzero leaves have been injected, the readout qubit has $\Pr(\text{ro}=1) \;=\; \sum_{\ell} w_\ell\,u_\ell$, the desired weighted average of branch values. For the Heat and Burgers’ stencils, we use two selector qubits $s_0,s_1$ and three nonzero leaves $00$, $01$, and $10$ (leaf $11$ has zero weight).
Fig.~\ref{fig:p5-leaf-injection} shows the circuits that apply the leaf-addressed rotations. Controls on logical $0$ are again implemented with $X$-control-$X$ sandwiches.

\paragraph{Putting it together: branching stencil micro-kernel.} We now assemble the full quantum branching micro-kernel for the three-point Heat/Burgers' stencil. The stencil weights and branch values are
$$
(w_L,\,w_C,\,w_R) = (\lambda,\,1-2\lambda,\,\lambda), \qquad u_L = u_{i-1}^n,\;\; u_C = u_i^n,\;\; u_R = u_{i+1}^n,\;\; 0\le\lambda\le\tfrac12.
$$
Two selector qubits $(s_0,s_1)$ prepare a three-way categorical distribution over leaves $\{00,01,10\}$ (leaf $11$ unused). A readout qubit $\text{ro}$ collects the branch values via the leaf-addressed rotations. The success probability of the readout is
\begin{equation}
\Pr(\text{ro}=1) \;=\; w_L\,u_{i-1}^n \;+\; w_C\,u_i^n \;+\; w_R\,u_{i+1}^n \;=\; \lambda\,u_{i-1}^n + (1-2\lambda)\,u_i^n + \lambda\,u_{i+1}^n,
\label{eq:ftcs-meter}
\end{equation}
which equals the updated value $u_i^{n+1}$ for the FTCS stencil.
\begin{figure}[t]
  \centering
  \resizebox{\columnwidth}{!}{%
  \begin{quantikz}
    \lstick{$s_0:\ket{0}$} & \gate{R_y(\theta_0)}
                           & \gate{X} & \ctrl{1} & \gate{X}
                           & \gate{X} & \ctrl{2} & \gate{X}
                           & \gate{X} & \ctrl{2} & \gate{X}
                           & \ctrl{2}
                           & \qw \\
    \lstick{$s_1:\ket{0}$} & \qw
                           & \qw      & \gate{R_y(\theta_L)} & \qw
                           & \gate{X} & \ctrl{1} & \gate{X}
                           & \qw      & \ctrl{1} & \qw
                           & \qw
                           & \qw \\
    \lstick{$\text{ro}:\ket{0}$}
                           & \qw
                           & \qw      & \qw      & \qw
                           & \qw      & \gate{R_y(\phi_{L})} & \qw
                           & \qw      & \gate{R_y(\phi_{C})} & \qw
                           & \gate{R_y(\phi_{R})}
                           & \meter{} \\
  \end{quantikz}
  }%
  \caption{Quantum branching micro-kernel for the three-point stencil. The selectors $s_0,s_1$ implement the categorical weights $(w_L,w_C,w_R)$, and leaf-addressed rotations on the readout $\text{ro}$ encode the neighbor values $(u_{i-1}^n,u_i^n,u_{i+1}^n)$. Measuring $\text{ro}$ yields an estimator for the FTCS update in Eq.~\ref{eq:ftcs-meter}.}
  \label{fig:heat-stencil-meas}
\end{figure}
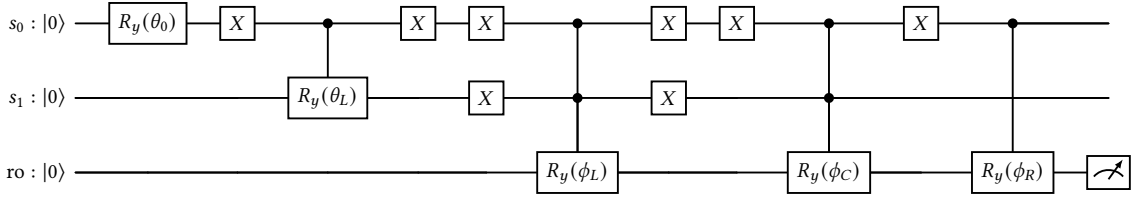

\begin{defbox}{}
\paragraph{Spatial sampling error.}
In the branching micro-kernel, the estimator is the sample mean of $U = u_X^n$ with $\mathbb{P}[X=b]=w_b$. Thus
$$
\mathrm{SE}\!\left(\hat u_i^{\,n+1}\right)
= \sqrt{\frac{\mathrm{Var}(U)}{M}}
= \sqrt{\frac{\sum_b w_b (u_b^{\,n})^2 - \bigl(\sum_b w_b u_b^{\,n}\bigr)^2}{M}}
\;\le\; \frac{\operatorname{range}\{u_b^{\,n}\}}{2\sqrt{M}}.
$$
This follows by writing $\mathrm{Var}(U)=\mathbb{E}[U^2]-\mathbb{E}[U]^2 =\sum_b w_b (u_b^{\,n})^2-\bigl(\sum_b w_b u_b^{\,n}\bigr)^2$ and then applying Popoviciu’s inequality $\mathrm{Var}(U)\le \tfrac{1}{4}\operatorname{range}\{u_b^{\,n}\}^2$~\cite{BhatiaDavis2000}, so $\mathrm{SE}(\hat u_i^{\,n+1})\le \operatorname{range}\{u_b^{\,n}\}/(2\sqrt{M})$. In our encoding $u_b^{\,n}\in[0,1]$, we have $\operatorname{range}\{u_b^{\,n}\}\le 1$ and therefore
$\mathrm{SE}(\hat u_i^{\,n+1}) \le 1/(2\sqrt{M})$. Error is largest where the stencil has high local variation (large spread among $u_{i-1}^n,u_i^n,u_{i+1}^n$, e.g., near steep gradients) and smallest in regions where neighbors are nearly equal.
\end{defbox}

\subsubsection{Additional QPU Micro-Kernels}
Beyond the basic Bernoulli and branching QPU, other shallow kernels can cover common stencil variants and linear-algebra subroutines. 

\paragraph{Signed-mixture micro-kernels.} When a local update involves signed coefficients (some positive, some negative), we use a signed-mixture construction. One option is a \ul{Bernoulli split}: we separate the positive and negative contributions into two convex sums, estimate each with a standard micro-kernel, and subtract the two results. A second option is a \ul{branching signed-mixture}. We sample a branch with probability proportional to the magnitude of its coefficient and keep track of its sign separately. In practice, this can be done either by encoding signed values directly on the readout or by encoding magnitudes and applying the sign at readout. For example, with a leaf-addressed phase flip or by negating the sample in software. Both variants yield unbiased estimates of updates with signed weights while reusing the same shallow quantum building blocks.

\paragraph{Row micro-kernel.} The discretization of elliptic PDEs leads to sparse linear systems $Au=b$. Iterative solvers such as Jacobi and Gauss-Seidel update each row $i$ by computing the off-diagonal contribution $\sum_{j\ne i} a_{ij} u_j$. The row micro-kernel estimates this quantity using a signed-mixture selector: it picks an index $j$ with probability proportional to $|a_{ij}|$, returns the corresponding value with the correct sign, and averages over many shots. This gives an unbiased estimate of the off-diagonal dot product, which we can plug into the usual update formula for $u_i$ in Jacobi or Gauss-Seidel.

\begin{figure}[t]
  \centering
  \resizebox{0.3\columnwidth}{!}{%
  \begin{quantikz}[row sep=0.22cm, column sep=0.6cm]
    \lstick{$\mathrm{ro}_1:\ket{0}$} & \gate{R_y(\pi/2)} & \meter{} \\
    \lstick{$\mathrm{ro}_2:\ket{0}$} & \gate{R_y(\pi/2)} & \meter{} \\
    \lstick{$\mathrm{ro}_3:\ket{0}$} & \gate{R_y(\pi/2)} & \meter{} \\
    \lstick{$\mathrm{ro}_4:\ket{0}$} & \gate{R_y(\pi/2)} & \meter{} \\
    \lstick{$\mathrm{ro}_5:\ket{0}$} & \gate{R_y(\pi/2)} & \meter{}
  \end{quantikz}}
  \caption{\ul{Noise micro-kernel} (Bernoulli coin-sum with $k_{\mathrm{coin}}{=}5$). Each qubit encodes a fair coin via $R_y(\pi/2)$. The measurement outcomes are
  combined in software to produce a Gaussian-like random increment.}
  \label{fig:coin-sum-noise}
\end{figure}
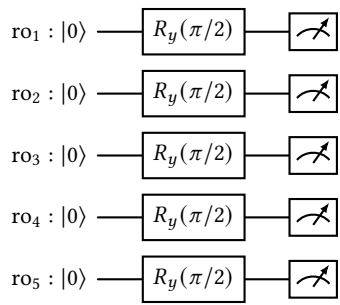

\paragraph{Noise micro-kernel.} Differently from the mapping kernels above, the noise micro-kernel is \emph{generative}. It returns either a random increment (added to the deterministic stencil at each node) or summary quantities of that increment for statistics-based solvers. Two realizations can be used. A small branching selector for finite laws, e.g., discrete Gaussian, Poisson jumps. Or a \emph{Bernoulli coin-sum} that approximates a zero-mean Gaussian by summing $k_{\mathrm{coin}}$ independent fair coins (see Fig.~\ref{fig:coin-sum-noise}). In the coin-sum variant, each coin is a one-qubit $R_y(\pi/2)$ encoder. The $k_{\mathrm{coin}}$ measurement bits are decoded to $\{\pm1\}$ and summed in software, then scaled to set the desired variance. Larger $k_{\mathrm{coin}}$ improves the Gaussian approximation. These kernels are needed when the model includes stochastic forcing (e.g., stochastic Heat and Burgers' equations, or PDEs with random sources).

\subsection{Extension to Higher-Dimensional and Higher-Order Stencils} 
The Bernoulli and branching micro-kernels used in 1D extend directly to $D$ dimensions and to higher-order stencils. Conceptually, nothing changes. At each node we still update the solution by taking a weighted average of a fixed set of neighboring values, with non-negative weights that sum to one (under the CFL conditions). In multiple dimensions, a standard second-order diffusion scheme uses the value at the center and its $2D$ axial neighbors. For example, in 2D this gives a 5-point stencil (center plus four directions), and in 3D a 7-point stencil. Higher-order schemes or upwind discretizations simply enlarge the neighborhood and increase the number of contributing points (e.g., 9-point in 2D or 27-point in 3D), but the update is still a convex combination of neighbors. When some coefficients become negative, the signed-mixture micro-kernels introduced earlier can be used instead.

For the Bernoulli micro-kernel, the extension is straightforward. Each neighbor value is first mapped into $[0,1]$, encoded on a single qubit with a Bernoulli encoder, and estimated from repeated measurements. Moving to higher dimensions or higher-order stencils only increases the number of such one-qubit encoders per node. The circuit structure and shot allocation remain unchanged.

For the branching micro-kernel, we need enough selector qubits to label all branches of the stencil. If there are $N_{\text{br}}$ branches, we choose the number of selector qubits $q_s$ so that $2^{q_s} \ge N_{\text{br}}$. A rotation tree on these selector qubits prepares a superposition whose amplitudes are
determined by the stencil weights. A single readout qubit receives leaf-addressed $R_y$ rotations that encode the corresponding neighbor values. For 5-point (2D) and 7-point (3D) diffusion stencils, three selector qubits are sufficient. Higher-order stencils increase $N_{\text{br}}$ and thus the number of selector states. The circuit depth grows with the rotation tree and the number of addressed injections, but the construction remains local and stencil-independent.

\section{Runtime Strategies: Batching and Fusion}
On current accelerators, including QPUs, the wall time for many small programs is often dominated by job submission/initialization, scheduling, and launch overheads~\cite{ekelund2025boosting,cicconetti2025modeling}. To amortize these fixed costs across many node updates, we use two complementary strategies: \textit{batched submission} and \textit{In-Circuit Fusion (ICF)}. Both return multiple node estimates per job, but they differ in circuit depth and qubit footprint.

In batched submission we package many small, independent circuits into a single job (for example, one Bernoulli encoder per node, or the three one-qubit encoders of a convex-sum node). The fixed overhead is then paid once while circuits remain shallow and unentangled. Each node still uses $M$ shots to reach a target precision, but host-to-device and job overheads are shared by all circuits in the batch.

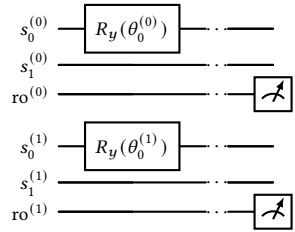
\begin{figure}[h!]
  \centering
  {\footnotesize
  \begin{quantikz}[row sep=0.15cm, column sep=0.35cm]
    \lstick{$s_0^{(0)}$}         & \gate{R_y(\theta_0^{(0)})} & \dots & \qw \\
    \lstick{$s_1^{(0)}$}         & \qw                        & \dots & \qw \\
    \lstick{$\mathrm{ro}^{(0)}$} & \qw                        & \dots & \meter{} \\
    \lstick{$s_0^{(1)}$}         & \gate{R_y(\theta_0^{(1)})} & \dots & \qw \\
    \lstick{$s_1^{(1)}$}         & \qw                        & \dots & \qw \\
    \lstick{$\mathrm{ro}^{(1)}$} & \qw                        & \dots & \meter{}
  \end{quantikz}}
  \caption{In-circuit fusion (diagram for $k_{\mathrm{fused}}{=}2$).}
  \label{fig:fused-2}
\end{figure}

In ICF, we place $k_{\mathrm{fused}}$ node micro-kernels side by side in a single circuit on disjoint qubit subsets. In this setup, one shot returns outcomes for all $k_{\mathrm{fused}}$ nodes. Thus $M$ shots leads to $M$ samples per node in the fused group, providing an approximate $k_{\mathrm{fused}}\times$ amortization of fixed overheads. The trade-off is resources. For the branching micro-kernel, ICF uses $3k_{\mathrm{fused}}$ qubits (two selectors plus one readout per node). While for the Bernoulli micro-kernel it uses $k_{\mathrm{fused}}$ qubits (one per node). More importantly, compiled depth typically grows with $k_{\mathrm{fused}}$ due to routing and control serialization, increasing two-qubit and readout errors on NISQ devices. Fig.~\ref{fig:fused-2} illustrates ICF schematically for $k_{\mathrm{fused}}{=}2$. 

A simple \ul{performance model} is useful to clarify the benefits of batching and ICF. If $T_{\mathrm{launch}}$ denotes fixed per-job overhead and $T_{\mathrm{node}}$ is the time contribution to advance one node in isolation (prep/transpile/execute/readout), then for ICF with $k_{\mathrm{fused}}$ nodes,
$
T_{\mathrm{per\text{-}node}}^{(\mathrm{ICF})} \;\approx\; \frac{T_{\mathrm{launch}}}{k_{\mathrm{fused}}} \;+\; \gamma\,T_{\mathrm{node}},
$
where $\gamma \ge 1$ is a multiplicative overhead factor introduced by ICF, accounting for depth inflation (e.g., additional SWAP routing) and readout serialization when measuring $k_{\mathrm{fused}}$ qubits per shot. For batched submission,
$
T_{\mathrm{per\text{-}node}}^{(\mathrm{batch})} \;\approx\; \frac{T_{\mathrm{launch}}}{k_{\mathrm{fused}}} \;+\; T_{\mathrm{node}},
$
since each node’s circuit remains shallow and isolated while overhead is shared across the batch. Batching and ICF can be combined, batch size and associated sampling budgets can be optimized to achieve high-throughput~\cite{markidis2025hpc}, similarly to GPU kernels on classical accelerators.

\section{Experimental Setup}
All experiments use Python~3.11, Qiskit~2.1, and the Qiskit Runtime primitives (\emph{Sampler~V2}). We validate the micro-kernel circuit logic, and check the error metrics using the \texttt{Qiskit Aer} state vector simulator via \emph{Sampler~V2}, then execute selected benchmarks on IBM Quantum hardware via \texttt{qiskit-ibm-runtime} with the preset pass manager at optimization level~1.
Unless otherwise stated, we use uniform 1D grids with $N=64$ interior nodes in the spatial domain, spanning from 0 to 1, $\nu = 1$ for Heat, and $\nu = 0.001$ for Burgers' equation. $\Delta t$ is calculated automatically to satisfy the CFL condition. In our simulation, we vary the final time $T$ and number of samples (shots) $M$. Accuracy is reported using the \emph{maximum-norm} ($(L_\infty$) error and the \emph{root-mean-square} ($L_2$) error. For each $M$, we perform five independent repetitions and report the sample mean and sample standard deviation of $L_\infty$ and $L_2$.

\paragraph{Hardware Backend.} We use the IBM \texttt{Brisbane} quantum computer with 127 qubits. A calibration snapshot obtained during our runs is summarized in Table~\ref{tab:brisbane-cal}.

\begin{table}[h!]
\centering
\caption{Backend and calibration snapshot for the IBM \texttt{Brisbane} quantum computer during the reported runs.}
\label{tab:brisbane-cal}
\begin{tabular}{l l}
\hline
Backend & IBM \texttt{Brisbane}, 127-qubit quantum computer \\
$T_1$ summary [\textmu s] & mean $=229.5$, median $=226.4$, min $=86.38$, max $=417.8$ \\
$T_2$ summary [\textmu s] & mean $=138.1$, median $=134.0$, min $=12.33$, max $=464.2$ \\
Readout assignment error $\epsilon$ & mean $=0.03139$, median $=0.0188$, min $=0.006592$, max $=0.2349$ \\
Readout confusion matrix $M$ & $\begin{bmatrix} 0.9040 & 0.0112 \\ 0.0960 & 0.9888 \end{bmatrix}$ \\
\hline
\end{tabular}
\end{table}

In addition, we check the impact of a single-qubit readout-assignment calibration on the readout qubit immediately before the hardware runs. We repeatedly prepare $\ket{0}$ and $\ket{1}$ and measure each state for a few thousand shots, obtaining a $2{\times}2$ confusion matrix $M$ whose entries give the probability of measuring outcome $i$ when state $j$ was prepared. We then use the inverse of this matrix to approximately debias the observed outcome probabilities. This is followed by clipping the corrected probabilities to $[0,1]$ and renormalization.

\section{Results}
In this section, we first assess accuracy on a noiseless simulator for the Heat and viscous Burgers' equations, reporting $L_\infty$ and $L_2$ errors versus analytic/numerical references as the shot budget $M$ varies. We then run the Bernoulli micro-kernel on the IBM \texttt{Brisbane} quantum computer and summarize how runtime scales with $M$.

\subsection{Micro-Kernel Accuracy and Convergence}
\begin{figure}[h!]
\centering
    \begin{minipage}{0.49\linewidth}
        \centering
        \includegraphics[width=\linewidth, clip = true, trim = 0 0 0 0cm]{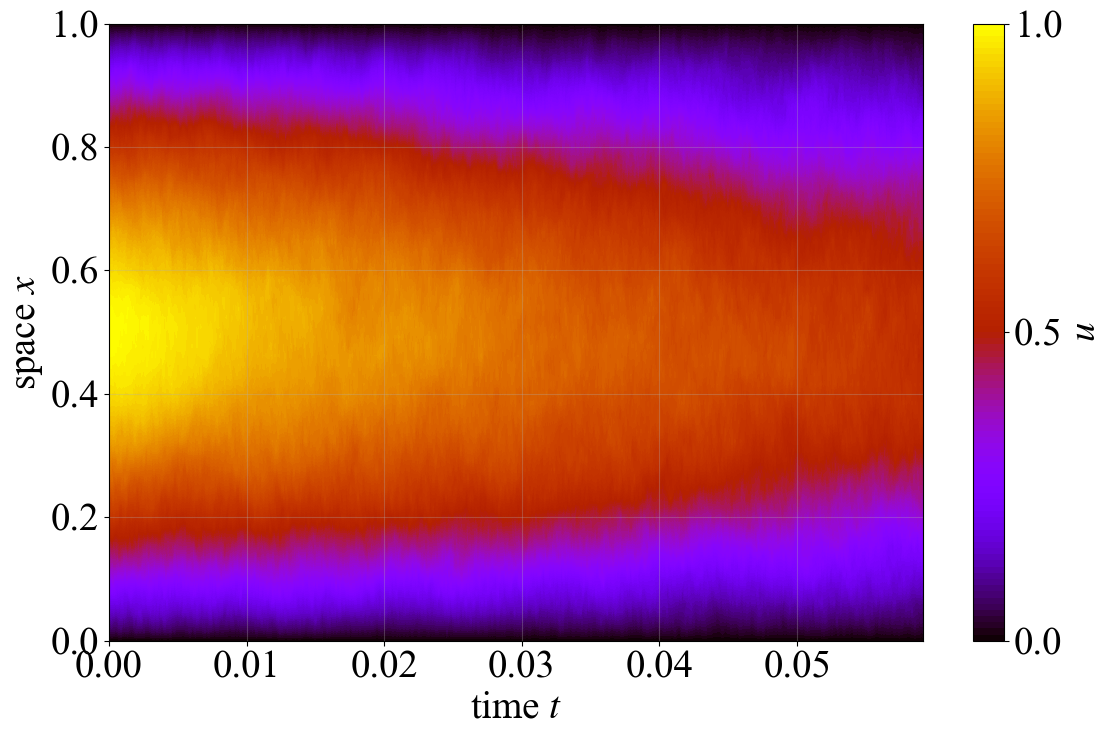}

        \vspace{-2mm}
        \caption*{(a) Heat equation.}
    \end{minipage}
    \begin{minipage}{0.49\linewidth}
        \centering
        \includegraphics[width=\linewidth, clip = true, trim = 0 0 0 0cm]{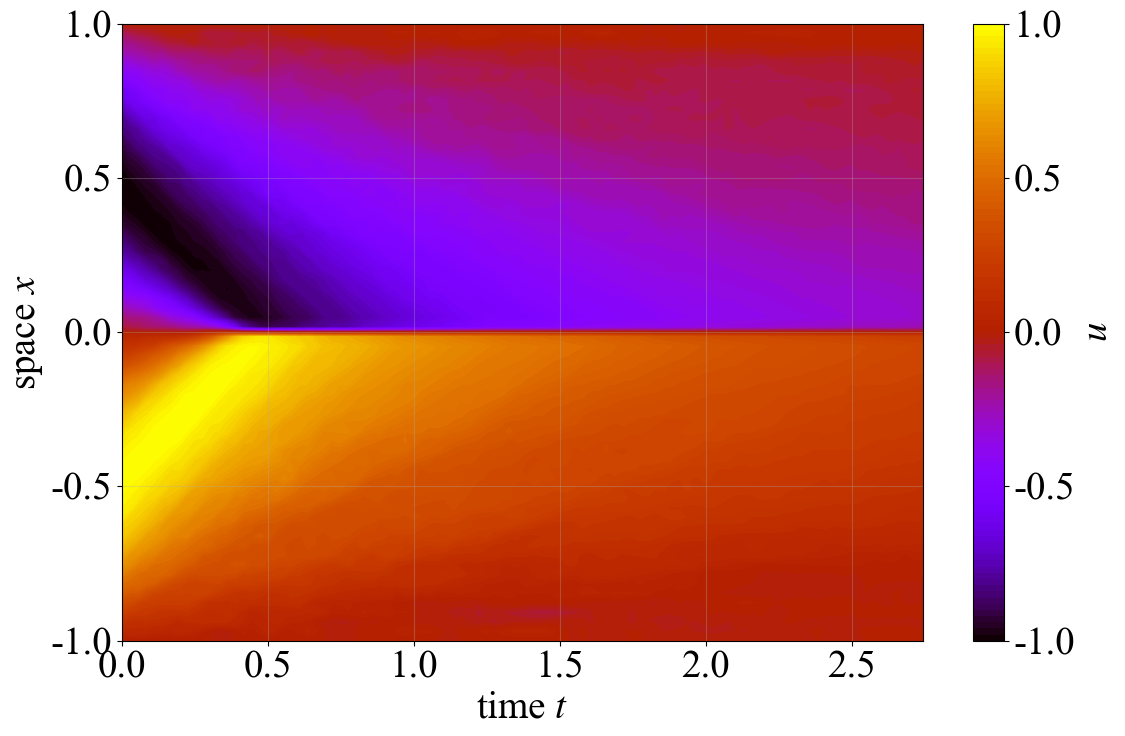}

        \vspace{-2mm}
        \caption*{(b) Burgers' equation.}
    \end{minipage}    
    \caption{Space-time plot of the solution evolution for the Heat and Burgers' equations, obtained with the branching micro-kernel on the noiseless \texttt{Qiskit Aer} quantum circuit simulator, with $M = 4{,}000$. (a) 1,000 time steps, $\nu = 1$; (b) 100 time steps, $\nu = 0.001$.}
    \label{fig:spacetime}
\end{figure}

As a first set of results, we validate the QPU micro-kernel method for the Heat and Burgers' equations against analytic and numerical solutions. In Fig.~\ref{fig:spacetime}, we show the space-time evolution of the solutions for the two PDEs, obtained with the branching micro-kernel on the noiseless \texttt{Qiskit Aer} quantum circuit simulator using $M = 4{,}000$ samples per grid point. In panel (a), the expected diffusive behavior from high to low $u$ values is visible in the solution of the Heat equation. The spatial error appears as a noisy transition region in areas with steep spatial gradients. Panel (b) shows the evolution for the advection-dominated Burgers' equation, with clear shock steepening.

\begin{figure}[h!]
\centering
    \begin{minipage}{0.45\linewidth}
        \centering
        \includegraphics[width=\linewidth, clip = true, trim = 0 0 0 1cm]{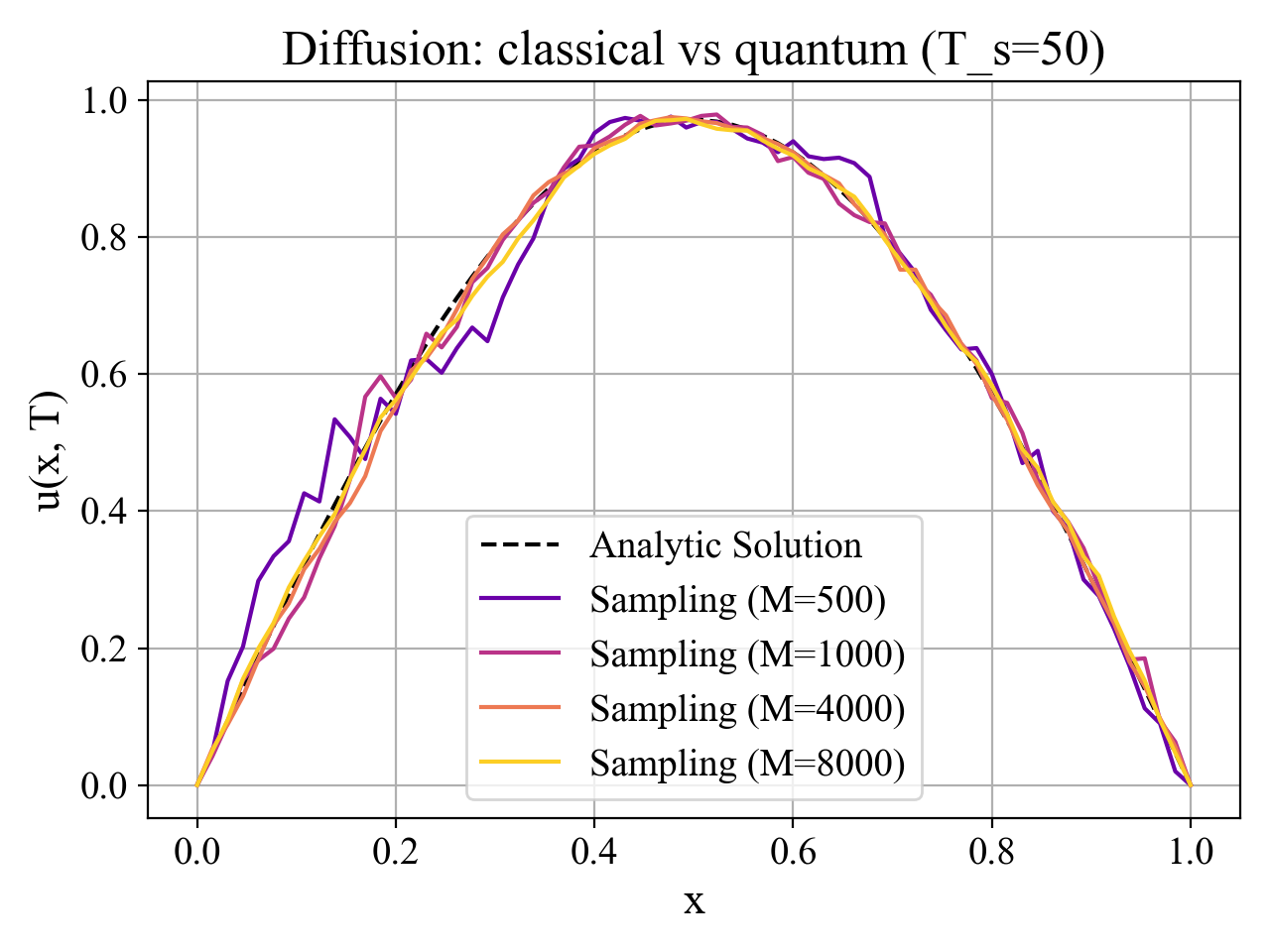}

        \vspace{-3mm}
        \caption*{(a) Heat equation.}
    \end{minipage}
    \begin{minipage}{0.45\linewidth}
        \centering
        \includegraphics[width=\linewidth, clip = true, trim = 0 0 0 1cm]{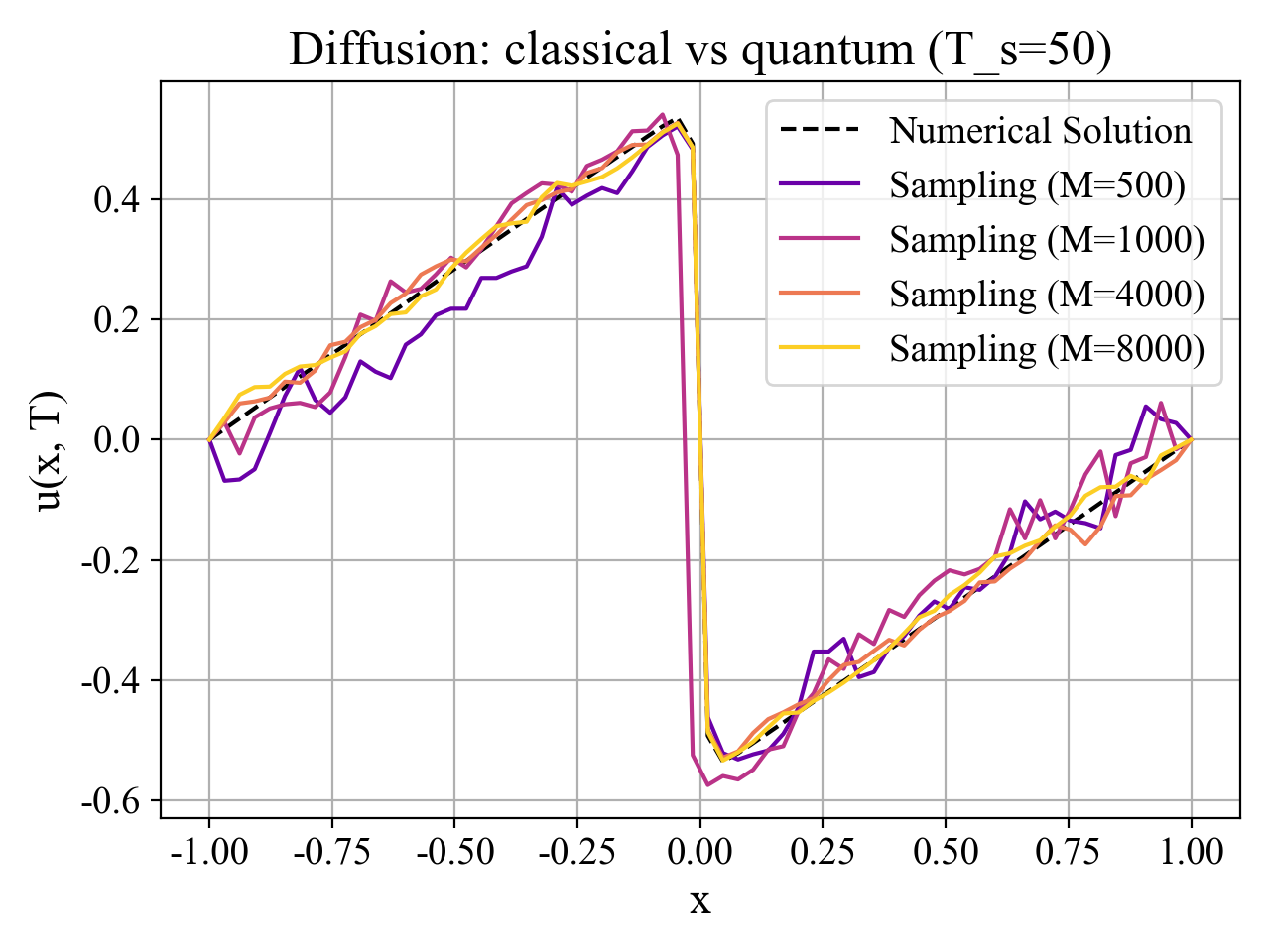}

        \vspace{-3mm}
        \caption*{(b) Burgers' equation.}
    \end{minipage}    
    \caption{Convergence of the shot-based solutions to the analytic reference as the number of shots $M$ increases, after 50 time steps. Higher number of shots yield higher accuracy.}
    \label{fig:convergence}
\end{figure}

As a second set of results, we show the impact of different numbers of samples using an implementation of the branching micro-kernel running on the noiseless \texttt{Qiskit Aer} quantum circuit simulator (so the only error source is statistical sample noise, not hardware noise). Fig.~\ref{fig:convergence} shows that, as we increase the number of samples from 500 to 8,000, the solution obtained with the micro-kernel converges toward the analytic and numerical solutions of the Heat (panel (a)) and Burgers' (panel (b)) equations.

\begin{figure}[h!]
    \centering
    
    \begin{minipage}{0.45\linewidth}
        \centering
        \includegraphics[width=\linewidth, clip = true, trim = 0 0 0 0]{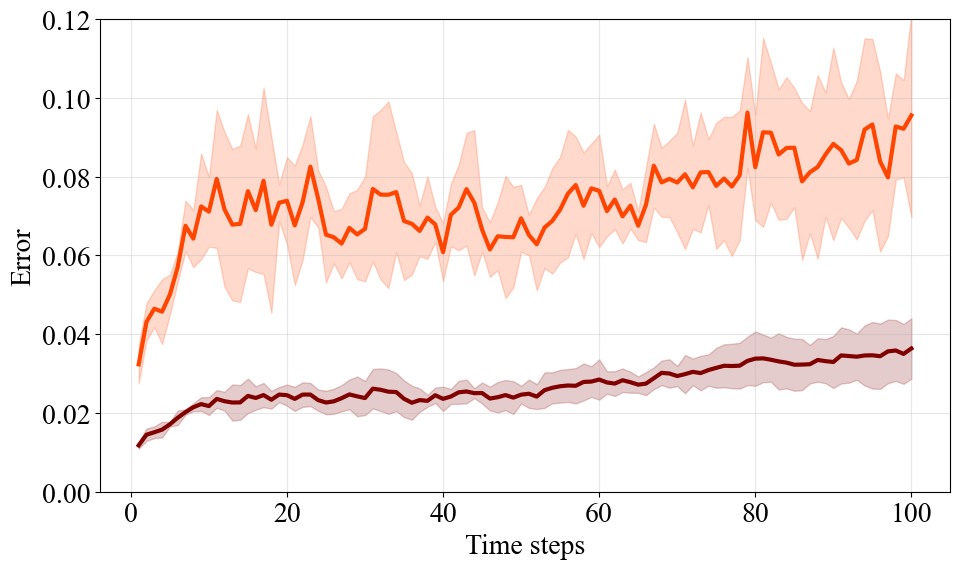}

        \vspace{-2mm}
        \caption*{(a) Heat equation, $M = 1{,}000$.}
    \end{minipage}
    \begin{minipage}{0.403\linewidth}
        \centering
        \includegraphics[width=\linewidth, clip = true, trim = 2.52cm 0 0 0]{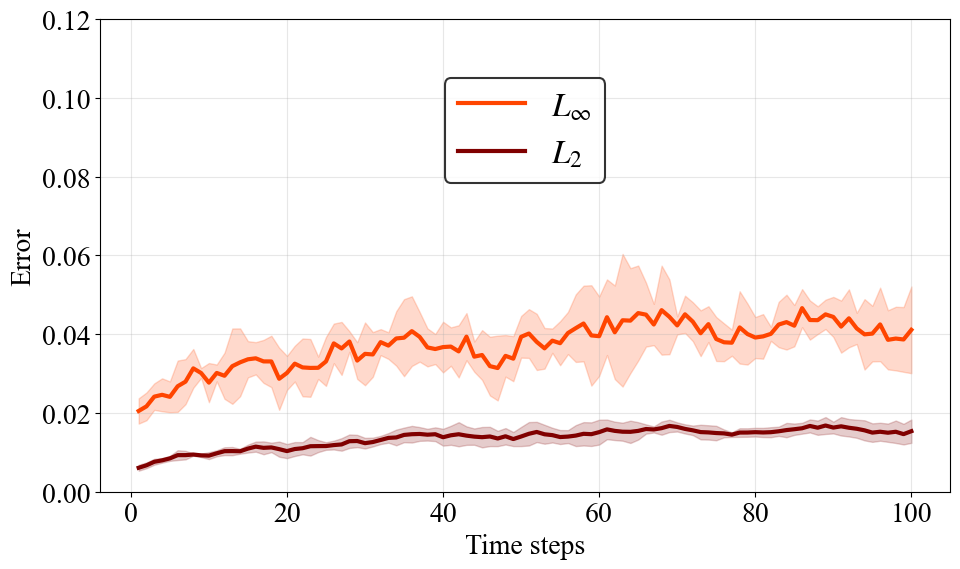}

        \vspace{-2mm}
        \caption*{(b) Heat equation, $M = 4{,}000$.}
    \end{minipage}

    \vspace{2mm}
    \begin{minipage}{0.45\linewidth}
        \centering
        \includegraphics[width=\linewidth, clip = true, trim = 0 0 0 0]{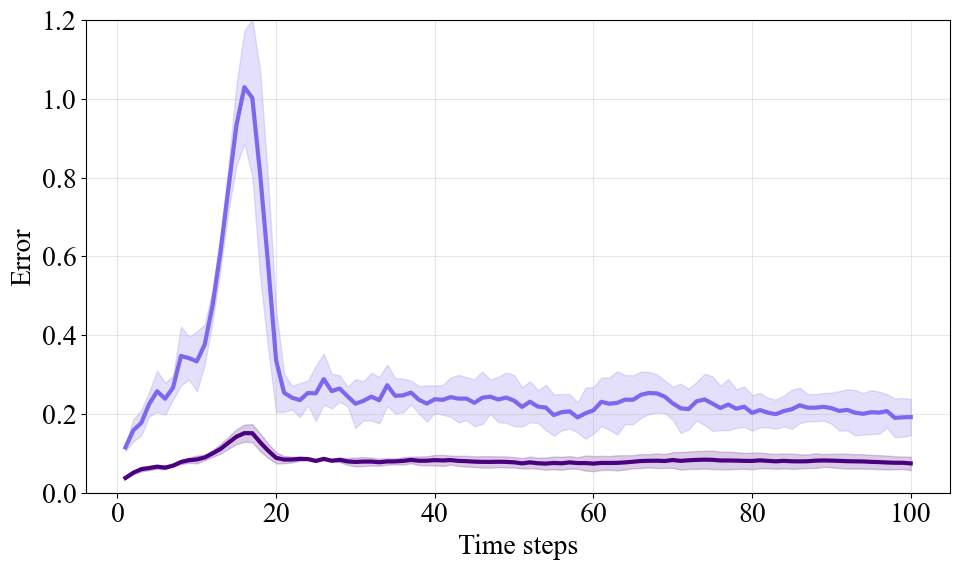}

        \vspace{-0mm}
        \caption*{(c) Burgers' equation, $M = 1{,}000$.}
    \end{minipage}
    \begin{minipage}{0.4\linewidth}
        \centering
        \includegraphics[width=\linewidth, clip = true, trim = 2.16cm 0 0 0]{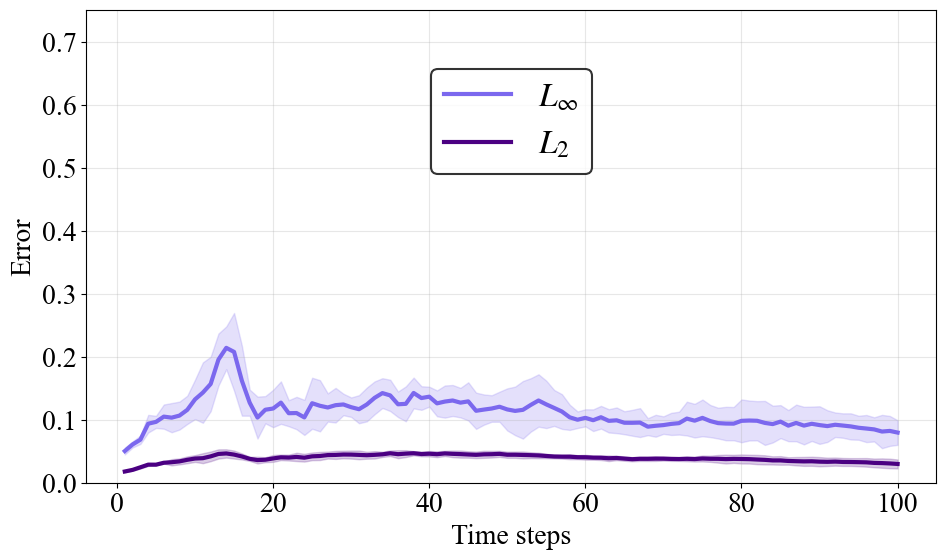}

        \vspace{-2mm}
        \caption*{(d) Burgers' equation, $M = 4{,}000$.}
    \end{minipage}
    \caption{Relative error ($L_2$, $L_\infty$) of the branching micro-kernel (with respect to the reference solution) for the Heat and Burgers' equations over 100 time steps, for different sample sizes $M$. Panels (a)-(b) show results for the Heat equation, and panels (c)-(d) for the Burgers' equation.}
    \label{fig:error-time}
\end{figure}

As a third set of results, we study error propagation over time using the branching micro-kernel and the noiseless \texttt{Qiskit Aer} quantum circuit simulator. Fig.~\ref{fig:error-time} shows the relative error ($L_2$ and $L_\infty$) over 100 time steps for the Heat (panels (a) and (b)) and Burgers' (panels (c) and (d)) equations, for two shot budgets ($M = 1{,}000$ and $M = 4{,}000$). The colored bands indicate the standard deviation, computed over five independent repetitions of the same simulation. Overall, we experimentally observe that the error remains bounded during the solution of both PDEs. After 100 time steps, the Heat equation exhibits an $L_2$ relative error of approximately $1.5\%$ and an $L_\infty$ relative error of about $4\%$ for $M = 4{,}000$, while the Burgers' equation shows about $7\%$ and $19\%$ for $L_2$ and $L_\infty$, respectively. As in Fig.~\ref{fig:convergence}, the noise decreases substantially as the number of samples increases.

We also note a clear peak in the error for the Burgers' equation (panels (c) and (d)). This peak is due to a delay in shock formation in the branching micro-kernel solution relative to the reference numerical solution. The solution obtained with the QPU branching micro-kernel develops the shock shortly afterward, leading to an accurate match of the $u$ profile after approximately 10 time steps. In our experiments, we found that the time-step offset required for the QPU micro-kernel solution to accurately capture the shock depends on the value of $\nu$.

\subsection{Evaluation of QPU Micro-Kernels on Quantum Hardware}
\label{sec:hw-results}
In addition to runs on the noiseless \texttt{Qiskit Aer} quantum simulator to assess the statistical noise of the QPU micro-kernel method, we also use the IBM \texttt{Brisbane} quantum computer to assess the potential of running QPU micro-kernels on current NISQ systems, without error correction. For these tests, we evaluate both the Bernoulli and branching micro-kernels for the Heat equation. We use a smaller grid ($N{=}15$ interior nodes) and perform only one iteration. Table~\ref{tab:circuit-profiles} presents the depths for the two micro-kernels in their basic formulation and after transpilation for the IBM \texttt{Brisbane} target, together with the number of one- and two-qubit gates in Brisbane’s native gate set.

\begin{table}[h!]
\centering
\caption{Circuit profiles for the kernels used in hardware runs (Qiskit 2.1, preset pass manager level~1; target basis \{\texttt{rz}, \texttt{sx}, \texttt{ecr}, \texttt{measure}\}).}
\label{tab:circuit-profiles}
\begin{tabular}{lrrrr}
\hline
\textbf{Kernel} & \textbf{Depth (untransp.)} & \textbf{Depth (transp.)} & \textbf{1q gates} & \textbf{2q gates} \\
\hline
Branching  & 12 & 118 & \(\texttt{rz}{:}80,\ \texttt{sx}{:}65,\ \texttt{x}{:}5\) & \(\texttt{ecr}{:}29\) \\
Bernoulli  &  2 &   3 & \(\texttt{rz}{:}1,\ \texttt{x}{:}1\)                     & 0 \\
\hline
\end{tabular}
\end{table}

\begin{figure}[h!]
  \centering
  \includegraphics[width=0.7\linewidth]{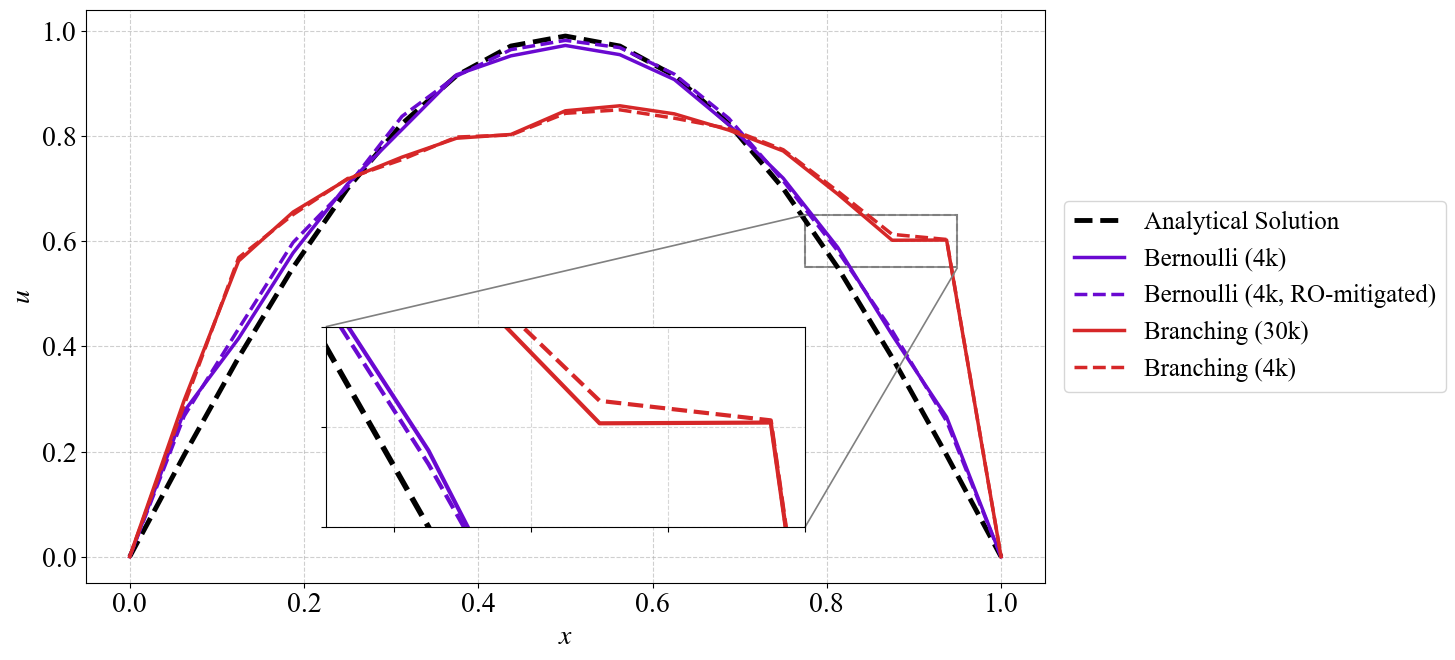}
  \caption{Heat equation (1D, IBM \texttt{Brisbane}). Single-step update on \texttt{ibm\_brisbane} ($N{=}15$). Analytic reference vs quantum estimates: Bernoulli ($M{=}4{,}000$, with/without 1-bit readout mitigation) and branching ($M{=}4{,}000$ and $M{=}30{,}000$). The dominant deviation from the analytic reference is due to hardware noise: the Bernoulli and branching solutions nearly overlap even though the number of samples differ.}
  \label{fig:ibm15_compare}
\end{figure}

On \texttt{Brisbane}, the branching kernel compiles to substantially deeper circuits ($12 \rightarrow 118$) with 29 two-qubit \texttt{ecr} gates. The circuit depth after transpilation limits the practical use of the branching micro-kernel on this device. In contrast, the Bernoulli kernel remains depth-minimal (depth $3$ after transpilation) with no two-qubit operations. In our tests, the Bernoulli micro-kernel provides acceptable performance, while the results with the branching micro-kernel are strongly impacted by device noise and decoherence. Fig.~\ref{fig:ibm15_compare} compares the solution obtained with Bernoulli (purple lines) and branching (red lines) micro-kernels with the analytic solution (black line).

For the Bernoulli micro-kernel, we use $M{=}4{,}000$ shots, with (dashed line) and without (solid line) readout error mitigation. The inset in Fig.~\ref{fig:ibm15_compare} shows that readout mitigation provides only a modest accuracy improvement in this setting. We also note that the error is mostly localized near mid-range $u$ values, in agreement with the spatial error analysis for the Bernoulli micro-kernel. For the branching micro-kernel, we perform two tests, with $M{=}4{,}000$ (dashed line) and $M{=}30{,}000$ (solid line) samples. In the latter case, the solution is significantly flattened and characterized by a large error. The number of samples has a negligible impact on the accuracy of the solution, indicating that noise from the quantum hardware is much larger than the statistical noise of the QPU micro-kernel method. Table~\ref{tab:ibm15_methods} summarizes the error values measured when running on IBM \texttt{Brisbane}.

\begin{table}[h!]
  \centering
  \caption{Diffusion (1D, hardware). Single-run errors vs analytic reference ($N{=}15$, one iteration).}
  \label{tab:ibm15_methods}
  \begin{tabular}{l r r}
    \hline
    \textbf{Method} & $L_\infty$ & $L_2$ \\
    \hline
    Bernoulli MK ($M{=}4{,}000$, no mit.)    & 0.0848 & 0.0368 \\
    Bernoulli MK ($M{=}4{,}000$, RO mit.)    & 0.0756 & 0.0378 \\
    Branching MK ($M{=}4{,}000$)             & 0.4116 & 0.1617 \\
    Branching MK ($M{=}30{,}000$)            & 0.4105 & 0.1592 \\
    \hline
  \end{tabular}
\end{table}

Finally, we collect telemetry data for the runs performed on the IBM \texttt{Brisbane} quantum computer, including execution times for transpilation, queuing, and circuit execution. For simplicity, we report telemetry for running the Bernoulli micro-kernel without batching or in-circuit fusion (ICF); this requires keeping an IBM Qiskit Runtime session open so that different runs share the same queue context. Using batching and ICF would further improve utilization of the quantum system. Table~\ref{tab:telemetry-nocal} reports the breakdown for the per-node Bernoulli micro-kernel with $M{=}4{,}000$ shots, and for the branching micro-kernel with $M{=}4{,}000$ and $M{=}30{,}000$ shots (no ICF). The number of jobs in the table corresponds to the number of QPU micro-kernels submitted, which is equal to the number of grid nodes (15) for the branching micro-kernel, and three times the number of grid nodes in the case of the Bernoulli micro-kernel.
\begin{table}[h!]
\centering
\caption{Timing comparison excluding readout-calibration jobs (when present). Mean $\pm$ standard deviation over per-node submissions.}
\label{tab:telemetry-nocal}
\begin{tabular}{l r c c c c}
\hline
\textbf{Config} & \textbf{Jobs} & \textbf{Transpile [s]} & \textbf{Queue [s]} & \textbf{Execution [s]} & \textbf{Wall [s]} \\
\hline
Branching 4k   & 15 & 0.022 $\pm$ 0.009 & 0.466 $\pm$ 0.131 & 3.501 $\pm$ 0.365 & 4.765 $\pm$ 0.404 \\
Branching 30k  & 15 & 0.029 $\pm$ 0.042 & 0.497 $\pm$ 0.101 & 11.421 $\pm$ 0.225 & 12.812 $\pm$ 0.353 \\
Bernoulli 4k   & 45 (${\mathrm{L}}, {\mathrm{C}}, {\mathrm{R}}$ $\times$ 15) & 0.005 $\pm$ 0.001 & 0.433 $\pm$ 0.119 & 3.485 $\pm$ 3.863 & 4.679 $\pm$ 3.903 \\
\hline
\end{tabular}
\end{table}
The largest computational cost is due to the actual circuit execution and sampling, with a maximum per-node execution time of about 11.4\,s for $M = 30{,}000$. When comparing the execution time for branching and Bernoulli at $M = 4{,}000$, we observe that circuit depth does not appear to significantly impact execution time. The queue time for the micro-kernels is approximately half a second, and transpilation time is negligible.

\section{Related Work}
\noindent \textbf{Monolithic Quantum Solvers.} Quantum computing offers several approaches to solving differential equations. One class of methods maps the PDE problem to simulating a Hamiltonian or solving a linear system. For example, the Quantum Linear Systems Algorithms (QLSAs) methods are quantum algorithms solving linear ODEs by encoding the time-evolution operator into a sparse linear system~\cite{berry2017quantum}. This approach, similar to the HHL algorithm, produces a quantum state proportional to the PDE solution at a final time. Similarly, Ref.~\cite{costa2019quantum} presents a quantum algorithm to simulate the wave equation by combining Hamiltonian simulation with QLSA techniques. While these algorithms promise asymptotic speedups, they typically assume fault-tolerant quantum hardware or long circuit depths. As a result, their direct applicability to NISQ devices is still limited.

\noindent \textbf{Hybrid Quantum-Classical Solvers.} To address NISQ limitations, variational and hybrid quantum-classical methods have been proposed for solving differential equations and their discretization~\cite{bravo2019variational}. Variational Quantum Algorithms (VQAs) use shallow parameterized circuits to represent the solution and a classical optimizer to train those parameters~\cite{cerezo2021variational}.  Variational Quantum Evolution Equation Solver are introduced for solving time-dependent PDEs~\cite{leong2022variational}. This method applies implicit time-stepping on a quantum ansatz state, and it reuses the solution from the previous time step as an encoded source to guide the next step’s optimization. A variational quantum method, based on imaginary-time evolution (VarQITE~\cite{mcardle2019variational}), to solve the Feynman-Kac PDE, arising from stochastic processes, is presented and discussed in Ref.~\cite{alghassi2022variational}. In addition to standard VQAs, differentiable quantum circuits have been used to solve nonlinear differential equations by defining a physics-based loss and optimizing circuit parameters via automatic differentiation through the quantum circuit~\cite{kyriienko2021solving}. Along the same line, Physics-Informed Neural-Network PINN) approaches have been ported to quantum hardware. In this case, parameterized quantum circuits act as neural surrogates trained to satisfy PDE constraints in a PINN-like fashion~\cite{markidis2022physics,panichi2025quantum}. All these variational approaches are promising for NISQ devices because they involve relatively shallow circuits and can be robust to noise. However, they often face challenges in scaling up, due to barren plateaus in optimization~\cite{mcclean2018barren} and the overhead of many circuit evaluations.

\noindent \textbf{Quantum Monte Carlo Methods.} A set of quantum algorithms, more similar to this work, for differential equations leverages Monte Carlo sampling and quantum amplitude estimation to obtain quadratic speedups. Many PDEs, especially parabolic equations, such as the heat equation or financial PDEs like Black-Scholes, can be solved via probabilistic interpretations that involve averaging over random paths. Quantum Monte Carlo algorithms aim to speed up these sampling tasks. Ref.~\cite{montanaro2015quantum} describes a general quantum algorithm to estimate the expected value of a random variable with bounded variance, achieving a quadratic speedup over classical Monte Carlo. The key subroutine is Quantum Amplitude Estimation (QAE), which uses phase-estimation techniques to quadratically reduce the number of samples needed for a given confidence level~\cite{brassard2000quantum}. In the context of PDEs, this implies quantum computers can in principle evaluate high-dimensional integrals or stochastic path averages more efficiently. Recent research has refined amplitude estimation to be more NISQ-friendly. For example, iterative or maximum-likelihood amplitude estimation algorithms~\cite{grinko2021iterative,suzuki2020amplitude,tanaka2021amplitude} eliminate the large-depth quantum Fourier transform of the original QAE. Quantum PDE solvers using QAE with Chebyshev integration points can improve accuracy and stability~\cite{oz2023efficient}. In this work, the use Chebyshev points can reduce the number of amplitude estimation queries needed by focusing on optimal integration nodes. There are also specialized Monte Carlo-inspired quantum algorithms. For instance, quantum-accelerated multilevel Monte Carlo methods have been proposed for stochastic PDEs in finance~\cite{an2021quantum,rebentrost2018quantum}, combining coarse and fine simulations with quantum speedups at each level to solve problems, such as option pricing more efficiently. 

\section{Discussion \& Conclusion}
In this work, we introduced QPU micro-kernels for stencil computations in explicit PDE solvers. Each node update is obtained by Monte Carlo sampling of a Bernoulli-encoded value, making the QPU act as a sampling accelerator. We applied the approach to two PDEs: the parabolic Heat equation and the nonlinear viscous Burgers’ equation. A quantum branching micro-kernel implements stencil weights within a superposition and drives multi-controlled rotations on a readout qubit. By contrast, a quantum Bernoulli micro-kernel is simpler and replaces in-circuit weighting by a classical mixture of up to three single-qubit circuits, with shot allocation proportional to the stencil weights. On ideal noiseless simulations, both realizations reproduce the expected update statistics. However, on current quantum hardware the branching variant suffers from increased depth and two-qubit gate errors after transpilation. This leads to a systematic downward bias in the measured probability. The Bernoulli variant, instead, eliminates entangling gates and confines the dominant error channel to single-qubit preparation and readout. In a representative run on IBM \texttt{Brisbane} quantum computer for a single explicit update with $N{=}15$ and $M{=}4{,}000$ shots per node, the Bernoulli micro-kernel produced acceptable accuracy relative to the analytical reference.

A practical limitation of the present study is that all reported estimates rely on direct sampling of a Bernoulli amplitude, and no variance reduction technique has been used. In principle, we can improve this to $O(1/M)$ by using \emph{Iterative Amplitude Estimation (IAE)}~\cite{grinko2021iterative,suzuki2020amplitude,tanaka2021amplitude}. Using IAE can lead to a near-quadratic reduction in the number of kernel invocations required to reach a given precision. However, each IAE round uses multiple coherent applications of the kernel and its inverse plus controlled reflections, which, for our branching and Bernoulli kernels, introduces entangling gates and depths that exceed current NISQ limitations~\cite{preskill2017quantum}. Moreover, IAE targets one scalar amplitude at a time. Running independent IAE loops for many grid nodes preclude our micro-kernel ICF strategy.

We note that the sampling error is spatially non-uniform. For the Bernoulli micro-kernel it is largest where the field is close to solution mid-values. For the branching micro-kernel, it tracks the local spread of stencil values and grows near steep gradients. These patterns persist over time and motivate an adaptive number of shots. We can use the previous time level as a pilot to guide allocation, giving more shots to nodes with small magnitudes or large neighbor variability, and fewer shots elsewhere, then renormalizing to a global measurement budget. In practice, we adopt an \emph{importance-sampling} variance-reduction~\cite{caflisch1998monte,robert1999monte} both within a node and across nodes. The orchestrator can increase sampling on heavily weighted or high-variance contributions and reduce it on negligible ones. 

It is important to note that a single micro-kernel execution on a near-term QPU does not deliver quantum advantage. A stencil update for the Heat equation requires $O(1)$ floating-point operations classically but $O(10^3$-$10^4)$ shots per node on present hardware to match similar precision. That said, the micro-kernel abstraction provides advantages. First, kernels are embarrassingly parallel across nodes and time steps, and are amenable to batching to amortize submission and queuing overheads. Second, kernel fusion (aggregating multiple nodes per circuit) can reduce launch costs when depth budgets allow. Third, improvements in measurement speed, classical-quantum orchestration, compilation, and qubit counts will enlarge the operating region where such sampling kernels become competitively useful as parallel sampling accelerators. Compared to alternative encodings of PDE updates, the proposed method offers a direct mapping from stencil coefficients to rotation angles and uses a small qubit footprint. 

The micro-kernels are most suited to parabolic and hyperbolic problems, where information propagates locally. When we applied similar ideas to elliptic problems (e.g., Poisson's equation) by embedding Jacobi, Gauss-Seidel, or relaxation schemes, convergence proved slow. This is because the update relies on global coupling and instantaneous propagation of boundary information. For such globally coupled systems, frequency-space or globally mixing kernels (e.g., QFT-based constructions) may be more appropriate than purely local, sampling-only micro-kernels.

Because the method is inherently Monte Carlo, it extends to high-dimensional integro-differential operators whose updates are nonlocal averages or collision integrals (radiative transfer, neutron transport, and kinetic plasma models with collisional terms). In these settings we compose micro-kernels under a classical orchestrator. We use Bernoulli and branching kernels for convex averages, and a noise kernel to inject stochastic increments when the PDE is stochastic. The orchestrator assigns shots and batches thousands of small jobs to meet runtime and accuracy requirements. For instance, radiation diffusion with random opacity~\cite{mihalas1984foundations} can use a Bernoulli kernel for the variable-coefficient diffusion blend and a noise kernel to add per-cell fluctuations. In multigroup neutron transport, scattering and fission sources can be built from angular/group fluxes~\cite{lewis1984computational} via Bernoulli encoders, while a branching kernel can sample absorption, scattering, and fission events in Monte Carlo tallies. We see QPU micro-kernels as a building block that scales across nodes and can benefit from NISQ hardware and runtime improvements.

\begin{acks}
Funded by the European Union. This work has received funding from the European High Performance Computing Joint Undertaking (JU) and Sweden, Finland, Germany, Greece, France, Slovenia, Spain, and Czech Republic under grant agreement No 101093261. 
\end{acks}

\bibliographystyle{ACM-Reference-Format}
\bibliography{QuantumMicroKernels}

\end{document}